\documentclass[aps,prb,twocolumn]{revtex4}
\usepackage{graphicx}
\newcommand{\abs}[1]{\vert #1 \vert}

\begin{document}
\title{Stripes in thin ferromagnetic films with out-of-plane anisotropy}
\author{D. Clarke}
\author{O. A. Tretiakov}
\author{O. Tchernyshyov}
\affiliation{Johns Hopkins University, Department of Physics and
Astronomy, 3400 N. Charles St., Baltimore, Maryland 21218}

\begin{abstract}
    We examine the $T=0$ phase diagram of a thin ferromagnetic film
    with a strong out-of-plane anisotropy (e.g. Co/Pt multilayers)
    in the vicinity of the reorientation phase transition.  The
    phase diagram in the anisotropy-applied field plane is universal
    in the limit in which the film thickness is the shortest length
    scale.  It contains uniform fully magnetized and canted phases,
    as well as periodically nonuniform states: weakly modulated
    spin-density waves and strongly modulated stripes.  We determine
    the boundaries of metastability of these phases and point out
    the existence of a critical point at which the difference
    between the SDW and striped phases vanishes. Out-of-plane
    magnetization curves exhibit hysteresis loops caused by the
    coexistence of one or more phases. Additionally, we study the
    effect of a system edge on the orientation of nearby
    stripes. We compare our results with recent experiments.
\end{abstract}
\maketitle
\section{Introduction}

    Recent experimental studies have revealed rich physics of thin
    ferromagnetic films with an easy axis of magnetization normal to
    the film plane.\cite{Davies04, Steren06, Schulte95, Donzelli03,
    Marty00, Shono00, Xiang05, Cheng06} Such films possess several
    magnetic phases in which the magnetization can be uniformly normal
    to the plane, canted, or periodically modulated in one direction
    (striped).  Magnetization curves exhibit intricate magnetic
    hysteresis indicating coexistence of phases.  Magnetic probes with
    sub-micron resolution provide detailed information about
    nucleation and growth of new domains during the magnetization
    reversal.\cite{Davies04} These developments fuel the need for a
    theoretical understanding of ferromagnetism in thin films.

    Theoretical studies of thin films with magnetic moments pointing out
    of the plane date back to the 1980s.\cite{Garel82}  The simplest
    model describes Ising spins with local ferromagnetic exchange and
    long-range dipolar interactions.  A competition between these forces
    makes the uniform ferromagnetic states unstable towards a spontaneous
    formation of magnetic stripes with an alternating sign of the
    magnetization.  The stripe period is a mesoscopic length determined
    by the relative strengths of exchange and dipolar interactions.

    A major drawback of the dipolar Ising model is the neglect of the
    in-plane components of magnetization, which become important when
    the magnetization rotates away from the plane normal, a process
    known as the reorientation phase transition (RPT).  A minimal model
    must therefore use a three-dimensional vector of magnetization whose
    magnitude $M$ is considered fixed well below the Curie temperature
    and whose orientation is given by the colatitude $\theta$ and
    azimuth $\phi$: $\mathbf M = M(\sin{\theta} \cos{\phi}, \,
    \sin{\theta} \sin{\phi}, \, \cos{\theta})$ with the $z$ axis normal
    to the film plane $xy$.  In a state with uniform magnetization, the
    energy density contains the easy-axis anisotropy $-K \cos^2{\theta}$
    and the demagnetizing term $(\mu_0/2) M^2\cos^2{\theta}$ due
    to the magnetic field. In the absence of an applied field the
    RPT occurs when the anisotropy drops below the critical strength
    $K_0 = \mu_0 M^2/2$.

    The assumption of uniform magnetization breaks down in the vicinity of
    the RPT: as in the dipolar Ising model, the competition between local
    and long-range forces leads to the formation of stripes with a
    mesoscopic period that depends on the effective anisotropy
    $K-K_0$, exchange strength $A$, and the film
    thickness $t$.\cite{Yafet88} Depending on the anisotropy and on the
    strength of an applied magnetic field, the stripes can appear as a
    weak spin-density wave (SDW) in the background of uniform
    magnetization or as fairly wide domains of uniform magnetization
    separated by narrow domain walls.

    The presence of the in-plane components of magnetization leads to an
    important distinction of domain walls from their counterparts in the
    Ising model: the vector of magnetization {\em rotates} between the
    (mostly) upward and downward directions.  Thus domain walls are endowed
    with in-plane magnetization, a fact with important topological
    consequences.  The domain walls are of the Bloch type: the in-plane
    magnetization on them points along the wall.  In contrast to Neel
    walls (in-plane magnetization normal to the wall), Bloch walls do not
    generate stray magnetic field and thus have a lower magnetic energy.
    We will therefore specialize to magnetization configurations in which
    the vector of magnetization depends on a single coordinate $x$ and lies
    in the $yz$ plane:
    \begin{equation}
        \mathbf M = M(0, \, \sin{\theta(x)}, \, \cos{\theta(x)}).
        \label{eq-Bloch}
    \end{equation}

    Using a variational approach, Berger and Erickson \cite{Berger97}
    obtained a phase diagram of such a one-dimensional model as a function
    of the anisotropy $K-K_0$ and an applied out-of-plane field $H_\perp$.
    It exhibits several phases with both first and second-order
    transitions between them.  Berger and Erickson focussed on {\em
    thermodynamic} transitions and did not provide boundaries of
    metastability. Such boundaries are important for the understanding of magnetic
    hysteresis in thin films. Often thermal activation is insufficient to
    initiate the decay of a metastable phase and magnetization
    reversal begins only when that phase becomes locally unstable.

    In this work we describe several new results.  Our main achievement is
    the determination of out-of-plane magnetization curves
    $M_\perp(H_\perp)$ that can be directly compared with experimental
    data.  We show that, for sufficiently thin films, the shape of the
    magnetization curve depends on a single parameter that is a function
    of the dimensionless effective anisotropy $\kappa = (K-K_0)/K_0$, film
    thickness $t$, and the magnetic exchange length $\lambda =
    \sqrt{2A/\mu_0 M^2}$.  This universality is the consequence of a
    scaling property of the free energy in the thin-film limit.  A proper
    rescaling of the anisotropy $\kappa$ and magnetic field $H_\perp$
    yields a universal phase diagram.  We point out the existence of
    critical points of the liquid-gas type that terminate lines of
    thermodynamic first-order phase transitions between striped and SDW
    phases. Finally, we discuss the behavior of stripes near the edge of the
    film. We find a tendency for stripes to meet an edge
    perpendicularly, independent of the particular shape of stripe.

    The formalism used in this work is minimization of the free energy.
    The average out-of-plane magnetization in a sample with stripes is chiefly dependent on the
    average period of those stripes rather than their orientational
    or translational order. We thus neglect the influence of thermal fluctuations, which tend to
    disrupt that orientational and translational order in the stripe
    phase.\cite{Kashuba93,Abanov95}

    The paper is organized as follows.  In Section \ref{model} we
    derive the functional of magnetic free energy specialized to
    one-dimensional variations, describe its scaling properties in a
    thin-film limit, and introduce appropriately rescaled variables.
    We illustrate its use by finding the lines of instability of the
    uniform phases in the anisotropy-applied field phase diagram.  In
    Section \ref{stripes} we determine the boundaries of
    metastability of nonuniform phases and magnetization curves
    using a numerical minimization.  We also find the location of
    the stripe-SDW critical points in the $(K, \, H_\perp)$ plane.
    Section \ref{edge} deals with the behavior of stripes near the
    film edge. Some useful intermediate results are described in the
    Appendixes.

\section{The model}
\label{model}

    \subsection{The free energy}

        The free-energy functional for a thin ferromagnetic film of
        thickness $t$ can be separated into a local and long-range parts.  The
        local part includes the exchange, uniaxial anisotropy, and
        Zeeman energies:
        \begin{equation}
            E_\mathrm{local}= t\int\mathrm{d}^2r \,
            \left(A|\nabla \hat \mathbf m|^2 - Km_z^2-\mu_0 M \,
            \mathbf H \cdot \hat \mathbf m\right),
            \label{eq-local}
        \end{equation}
        where $\hat \mathbf m = (\sin{\theta} \cos{\phi}, \, \sin{\theta}
        \sin{\phi}, \, \cos{\theta})$ is the three-dimensional unit vector
        pointing along the magnetization and $\nabla = (\partial_x, \, \partial_y)$
        is the two-dimensional gradient in the plane of the film.  The long-range
        part is due to dipolar interactions:
        \begin{equation}
            E_\mathrm{dipolar}=\frac{\mu_0M^2}{4\pi}
                \int \mathrm{d}^2r\int \mathrm{d}^2r' \,
                m_z(\mathbf r)V(\mathbf{r-r'})
                m_z(\mathbf r'),
            \label{eq-dipolar}
        \end{equation}
        where the dipolar kernel $V(\mathbf r) = 1/r - 1/\sqrt{r^2 + t^2}$
        reflects the interaction of magnetic charges with densities $\pm
        M m_z(\mathbf r)$ induced on the top and bottom surfaces of the film.
        The expression for the dipolar energy (\ref{eq-dipolar}) is valid
        provided that there are no magnetic charges in the bulk of the film,
        i.e. $\partial_x m_x + \partial_y m_y = 0$.  This condition is
        compatible with Eq.~(\ref{eq-Bloch}), which describes a system with Bloch
        domain walls.  Domain walls of the Neel type generate additional
        dipolar terms.

        Specializing to one-dimensional configurations without bulk magnetic
        charges (\ref{eq-Bloch}), and with magnetic field applied
        perpendicular to the plane of the film, we obtain the energy
        \begin{eqnarray}
            \frac{E}{L_y t} =\int\mathrm d x\left[
                A\left(\mathrm{d}\theta/\mathrm{d}x\right)^2
               -K\cos^2\theta
               -\mu_0 M H \cos{\theta})\right]
               \nonumber\\
               +\,\frac{1}{2}\mu_0 M^2\int\mathrm{d}x\int\mathrm{d}x' \,
               \cos{\theta(x)} \, V(x-x') \, \cos{\theta(x')},
               \label{fullenergy}
        \end{eqnarray}
        where $L_y$ is the width of the system in the $y$ direction.
        The one-dimensional kernel $V(x)=(1/2\pi t) \ln{(1+t^2/x^2)}$ obtained by
        integrating (\ref{eq-dipolar}) over $y$ has the
        Fourier transform
        \begin{equation}
            \tilde{V}(k)=\frac{1-e^{-\abs{k}t}}{\abs{k}t} = 1 - \frac{|k|t}{2}
                + \mathcal O(t^2).
            \label{momentumkernel}
        \end{equation}
        The Taylor expansion (\ref{momentumkernel}) is justified when the film
        thickness $t$ is the shortest length scale in the problem.  The
        zeroth-order term $\tilde{V}_0(k) = 1$ can be interpreted as a contact
        part of the dipolar interaction that simply shifts the anisotropy: $K
        \mapsto K-K_0$.  The first-order term $\tilde{V}_1(k) = -|k|t/2$
        represents the effect of the stray dipolar field.  Its inverse Fourier
        transform $V_1(x)$ diverges at short length scales and requires a
        short-range cutoff.  It has the following properties:
        \begin{equation}
            V_1(x) \sim \frac{t}{2\pi x^2} \mbox{ as } x \to \infty, \quad
            \int \mathrm d x \, V_1(x) = 0.
            \label{v1prop}
        \end{equation}

    \subsection{Scaling property of the free energy}
    \label{scaling}

        It is convenient to use natural scales for various physical units:
        $K_0 = \mu_0 M^2/2$ for the effective anisotropy $\kappa$ and
        volume energy density, and $M$ for magnetic field:
        \begin{equation}
            \kappa = (K-K_0)/K_0, \quad
            \mathbf h = \mathbf H/M.
        \end{equation}
        In these units,
        \begin{equation}
            \frac{E_\mathrm{local}}{\mu_0 M^2 L_y t} =  \int\mathrm d x\left[
               \frac{\lambda^2}{2} \left(\frac{\mathrm d \theta}{\mathrm d x}\right)^2
               -\frac{\kappa}{2}\cos^2\theta
               - h \cos{\theta}\right]
            \label{localE}
        \end{equation}
        and
        \begin{equation}
            \frac{E_\mathrm{stray}}{\mu_0 M^2 L_y t} = \frac{1}{2}
               \int\mathrm{d}x\int\mathrm{d}x' \,
               \cos{\theta(x)} \, V_1(x-x') \, \cos{\theta(x')},
            \label{strayE}
        \end{equation}
        where $\lambda = \sqrt{A/K_0}$ is the exchange length.

        The free energy, given by the sum of these terms, is invariant under a scaling
        transformation
        \begin{equation}
            x \mapsto bx, \quad
            t \mapsto t, \quad
            \lambda \mapsto b^{1/2} \lambda, \quad
            \kappa \mapsto b^{-1} \kappa, \quad
            \mathbf h \mapsto b^{-1} \mathbf h.
            \label{eq-scaling}
        \end{equation}
        This symmetry indicates that the state of the film depends on the
        effective anisotropy and magnetic field through scale-invariant
        variables $\kappa/\kappa_0$ and $\mathbf h/\kappa_0$, where $\kappa_0
        = t^2/(4\lambda)^2$ is an effective anisotropy scale whose
        significance will be clarified shortly.  It also yields a characteristic length
        scale $8 \pi \lambda^2/t$, which determines the period of the
        stripes---see Eq.~(\ref{eq-k0}) below.

        The free energy is scale-invariant only to the lowest order in the
        film thickness $t$.  Inclusion of higher-order terms in the dipolar
        kernel (\ref{momentumkernel}) violates this property.  The scaling
        works as long as the thickness $t$ is small compared to all other length
        scales, in particular the exchange length $\lambda \approx 5$ nm (for Co).

        Note that this scaling law applies more generally to
        two-dimensional configurations with an applied field in any
        direction, as the energies associated with bulk magnetic
        charges and in plane fields are also invariant under
        (\ref{eq-scaling}) when $t$ is small.

    \subsection{Uniform phases}
    \label{uniform-out}

        Let us first discuss the uniform phases and their instabilities in the
        case that the applied field is normal to the plane.  The case
        $\kappa>0$ is trivial: only the upward and downward polarized states
        are stable.  Both are locally stable in the region $|h|<\kappa$.
        The coexistence leads to a hysteresis in magnetization curves
        $m_z(h)$.

        The situation is more interesting for $\kappa<0$, where magnetization
        prefers the in-plane direction.  In a uniform state the stray field
        vanishes and the energy is proportional to $- (\kappa/2) \cos^2{\theta} -
        h \cos{\theta}$.  In a strong field, $|h|>|\kappa|$, the
        film is fully polarized, $\cos{\theta} = \mathrm{sgn}(h)$.
        Below the critical strength, $|h|<|\kappa|$, the magnetization
        is canted, $\cos{\theta} = h/|\kappa|$.  The uniform-to-canted
        transitions at $h=\pm\kappa$ are continuous.

        Consider the free energy of small fluctuations around a canted state,
        $\cos{\theta}(x) = -h/\kappa + \delta(x)$:
        \begin{equation}
            \Delta E = \frac{\mu_0 M^2 L_y t}{2}
                \int \frac{\mathrm{d}k}{2\pi}~\left(
                \frac{\lambda^2 k^2}{\sin^2{\theta_0}}-\kappa
                -\frac{\abs{k}t}{2}\right)|\tilde{\delta}(k)|^2 + \mathcal O(\delta^4),
        \end{equation}
        where $\tilde{\delta}(k)$ is the Fourier transform of $\delta(x)$ and
        $\cos{\theta_0} = -h/\kappa$ is the equilibrium value for the
        canted state.  The softest mode has the wavenumber
        \begin{equation}
            k_0 = \frac{t \sin^2{\theta_0}}{4\lambda^2}.
            \label{eq-k0}
        \end{equation}
        A spin-density wave develops in the canted background on the line
        \begin{equation}
            (h/\kappa)^2 = 1+ \kappa/\kappa_0,
            \label{eq-SDW}
        \end{equation}
        where $\kappa_0 = t^2/(4\lambda)^2$.  An expansion to the order $\mathcal
        O(\delta^4)$ reveals a positive-definite quartic term.  Thus the canted-SDW
        transition is also continuous.  Note that the scaling arguments are confirmed:
        the wavelength is indeed set by the scale $8\pi\lambda^2/t$ and the critical
        line (\ref{eq-SDW}) contains the rescaled variables $h/\kappa_0$
        and $\kappa/\kappa_0$.
        \begin{figure}[htbp]
            \begin{center}
                \includegraphics[width=\columnwidth]{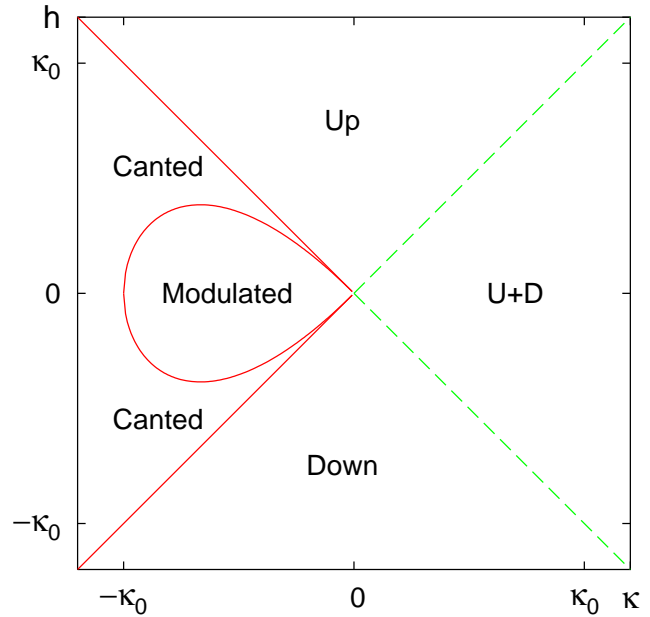}
            \end{center}
            \caption{(Color online) Phase diagram showing the regions of
                stability of uniform phases.
                Up (U) and Down (D) phases coexist in the region on the right.
                Solid and dashed lines denote
                continuous and discontinuous transitions, respectively.}
            \label{fig-uphases}
        \end{figure}

        Since no uniform solution is stable inside the semicubic parabola
        (\ref{eq-SDW}), this region must be occupied by an inhomogeneous
        state, which one might easily guess to be a stripe phase (see Fig.~\ref{fig-uphases}).  However, the
        situation is somewhat more complicated.  A further analysis shows that
        the stripe phase remains (at least locally) stable outside the
        semicubic curve.  In addition, we find a region of coexistence between
        a strongly inhomogeneous striped state and a weakly inhomogeneous SDW
        state.  Like a liquid and a gas, the two phases differ from each other
        only quantitatively.  Indeed, there is a critical point at which the
        differences vanish continuously, as we will discuss below.

\section{Stripe phase}\label{stripes}

    In this section, we describe the striped state in various regions of
    the $h-\kappa$ plane. We determine the approximate boundaries of
    stability of the striped phase, and show how the metastability of
    the stripe phase leads to the hysteresis curves observed in experiments.

    \subsection{Numerical determination of metastability
        limits for the stripe phase}\label{shortstripes}
        \begin{figure}[htbp]
            \begin{center}
                \includegraphics[width=\columnwidth]{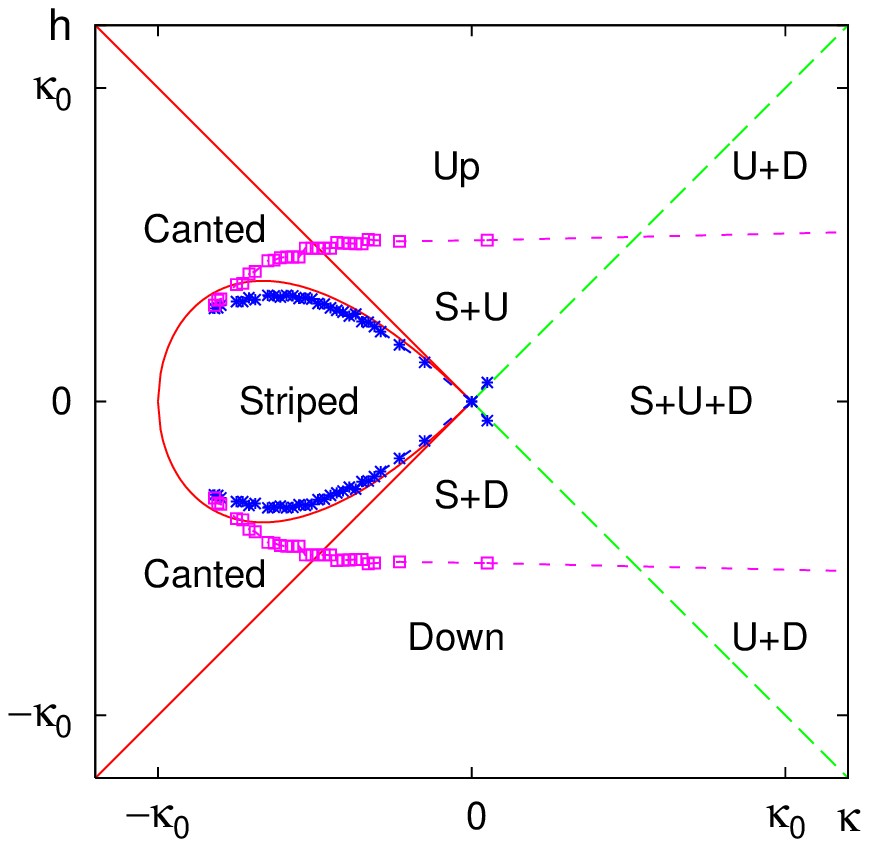}
                \includegraphics[width=\columnwidth]{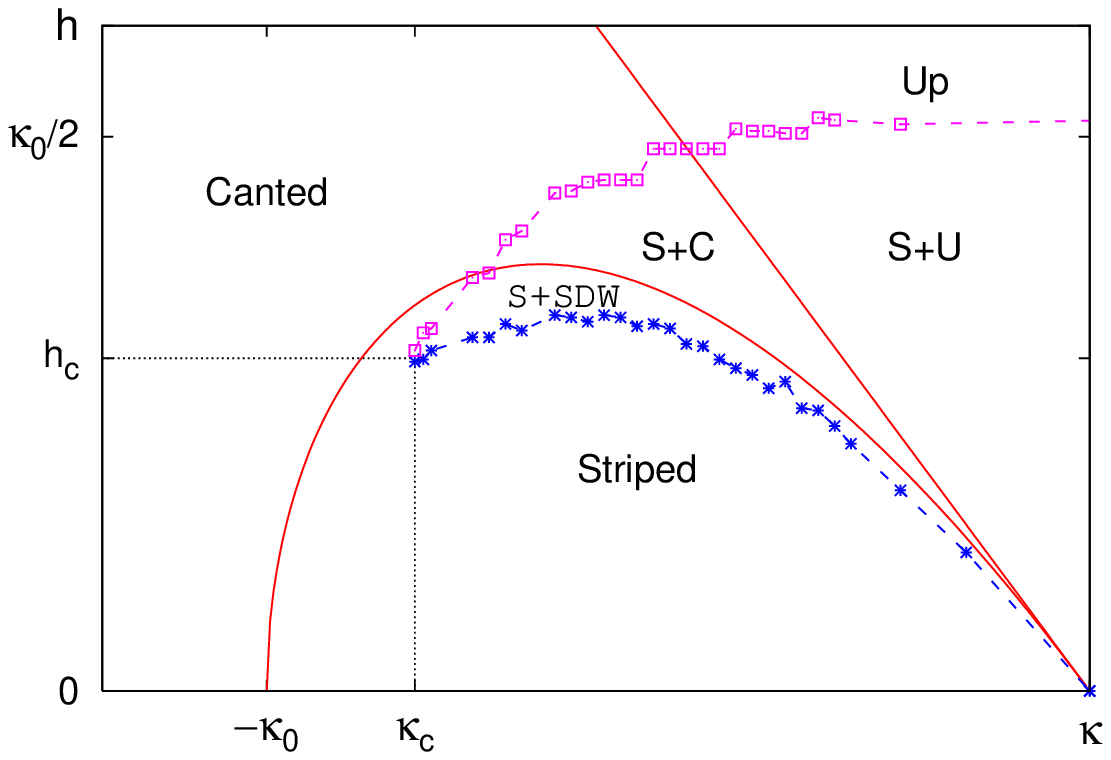}
            \end{center}
            \caption{(Color online)  Numerically determined boundaries
                of the nonuniform phases are shown as symbols.
                The Striped phase can coexist with the Canted (C),
                Up (U), Down (D), and Spin density wave (SDW) states
                in various regions of the phase diagram. Solid and dashed
                 lines denote continuous and discontinuous
                transitions, respectively.}
            \label{phase diagram}
        \end{figure}
\begin{figure}[htbp]
            \begin{center}
                \includegraphics[width=0.49\columnwidth]{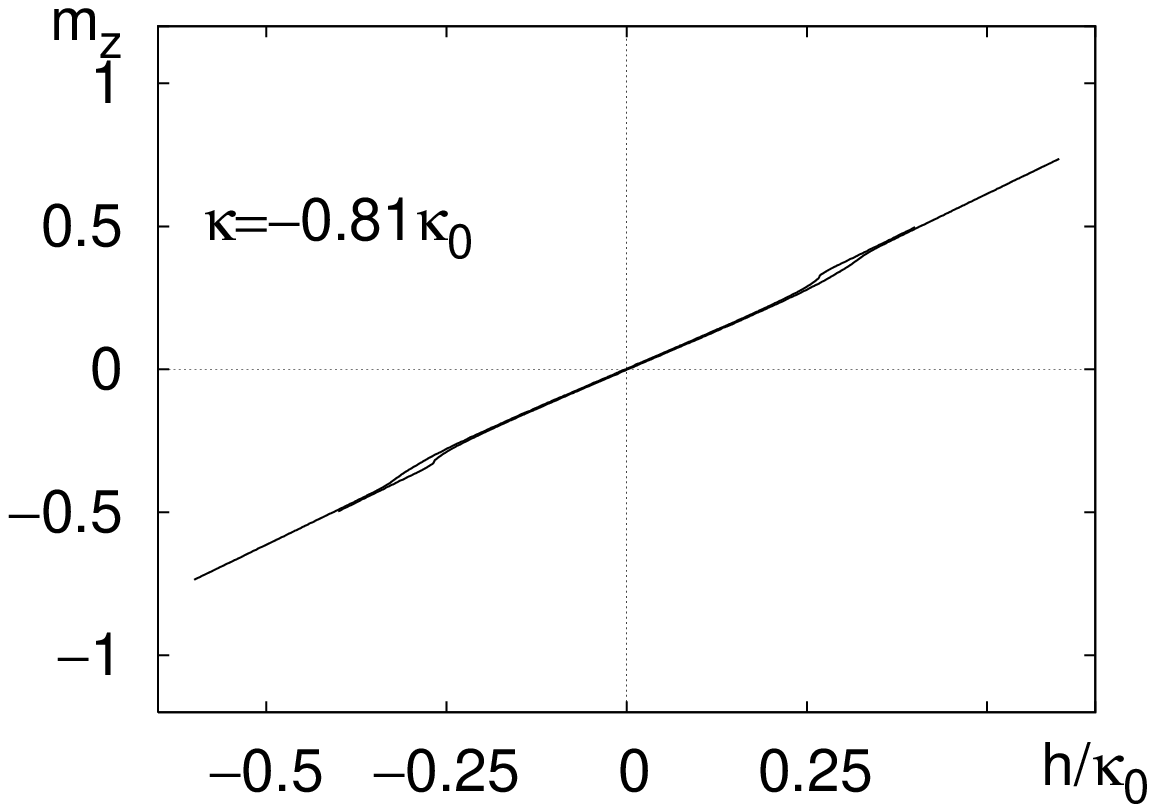}
                \includegraphics[width=0.49\columnwidth]{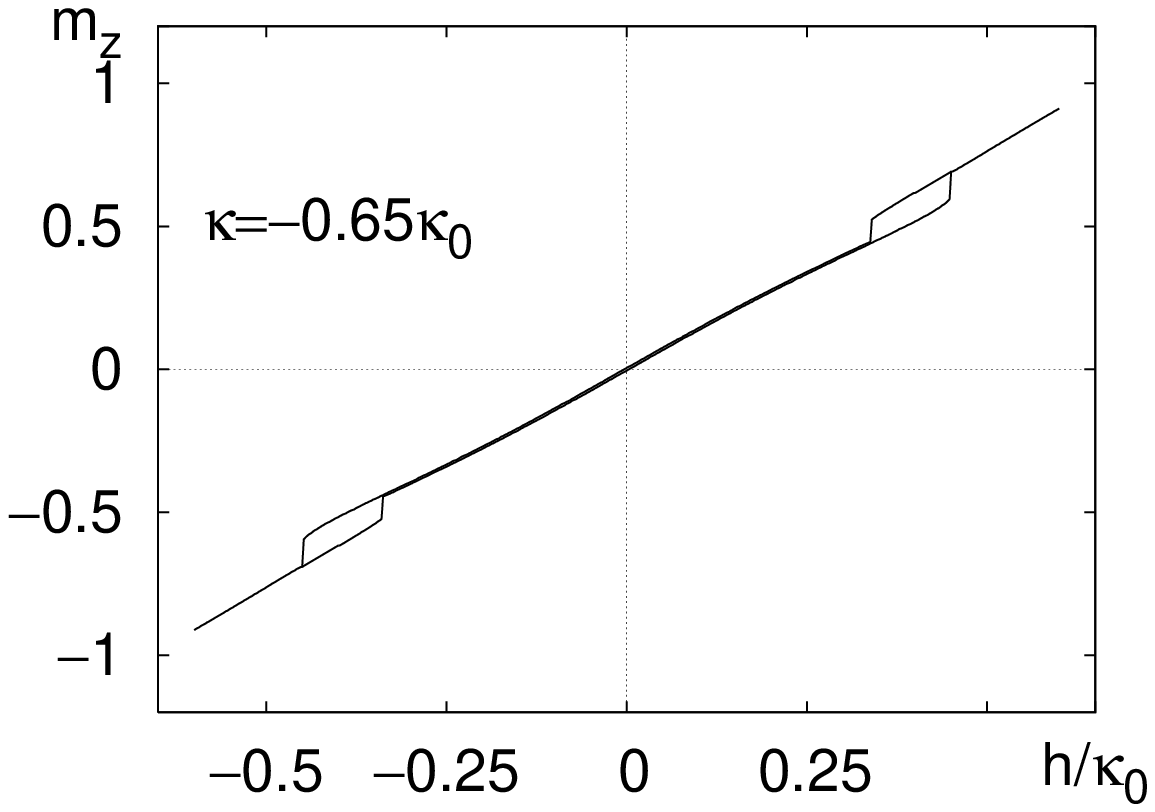}
                \includegraphics[width=0.49\columnwidth]{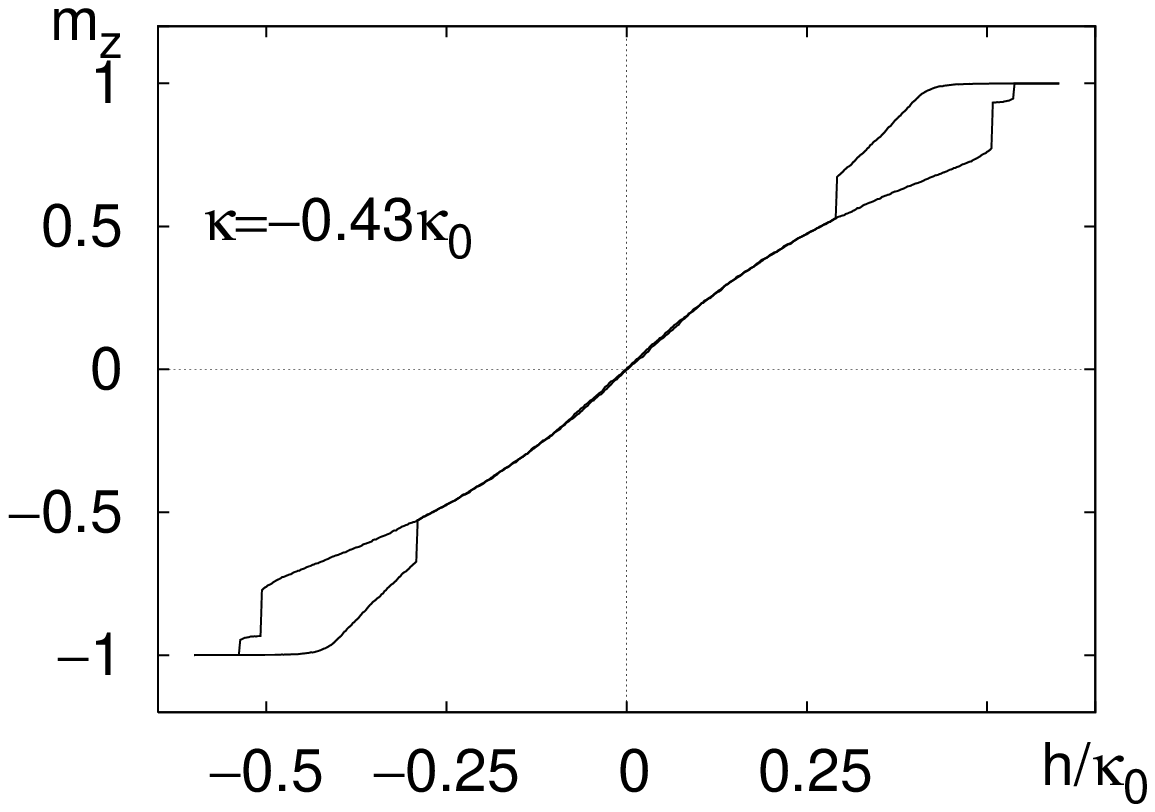}
                \includegraphics[width=0.49\columnwidth]{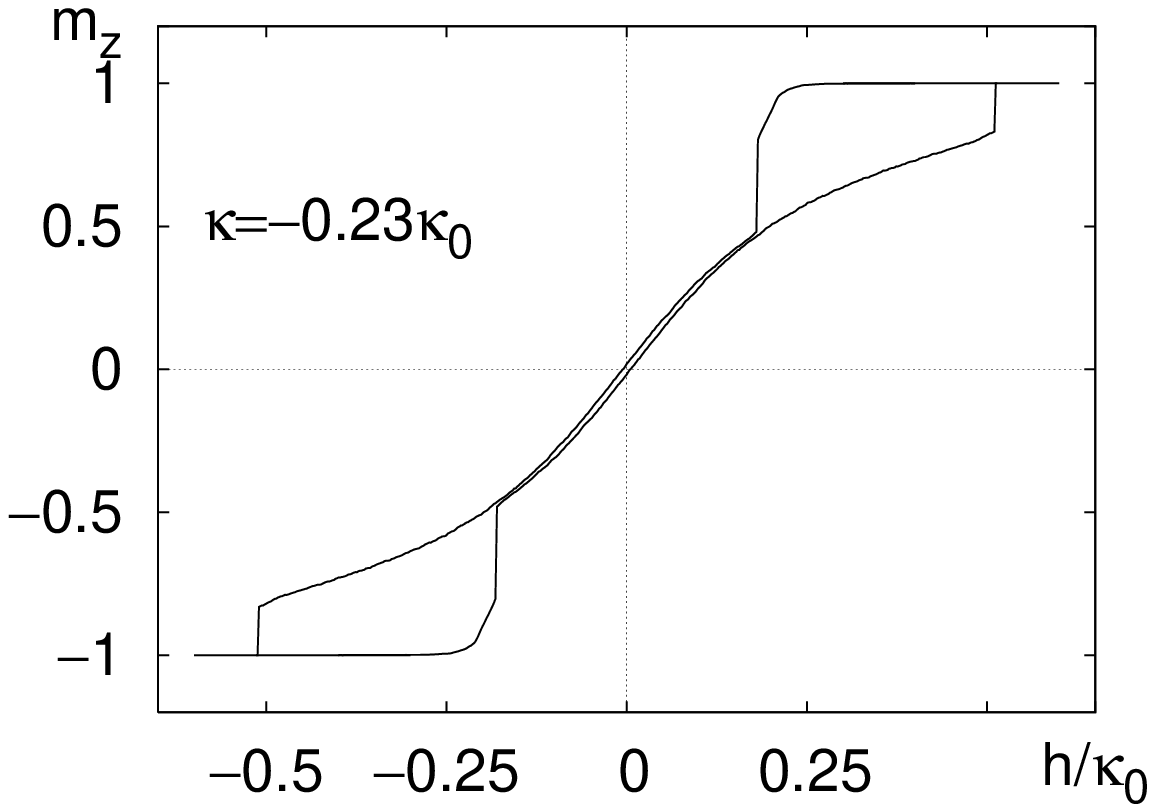}
                \includegraphics[width=0.49\columnwidth]{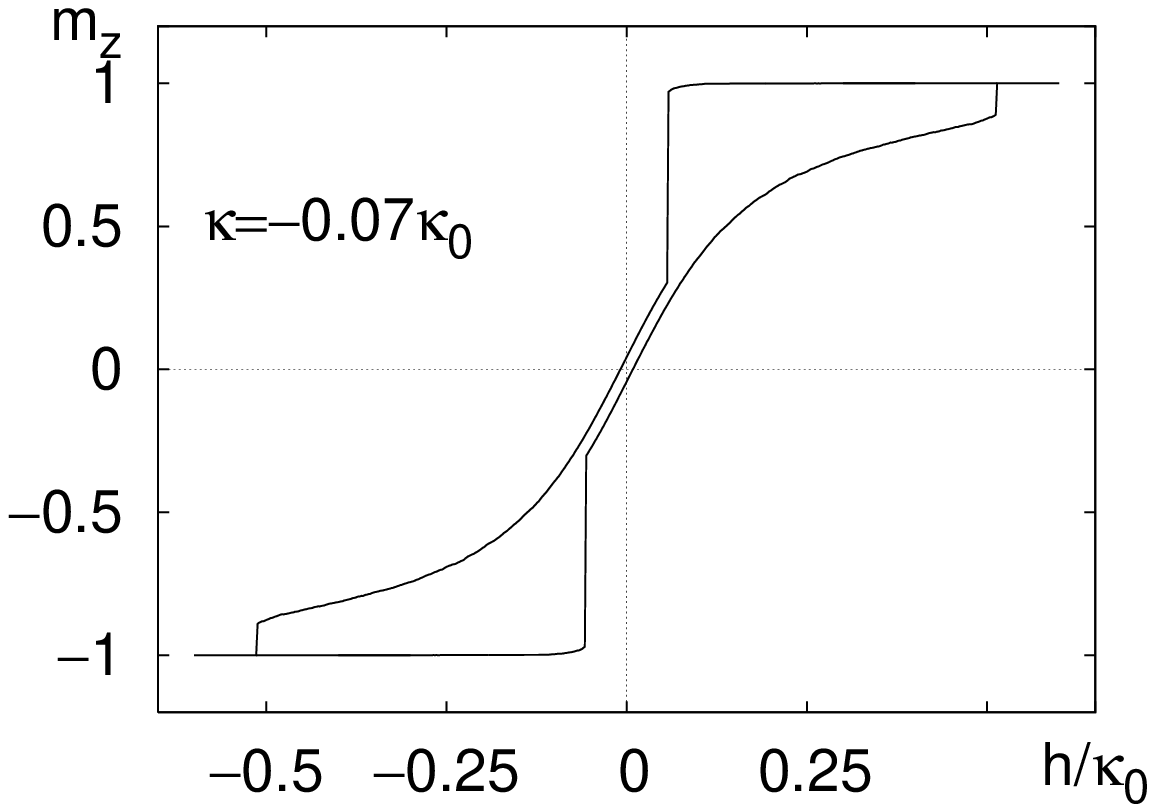}
                \includegraphics[width=0.49\columnwidth]{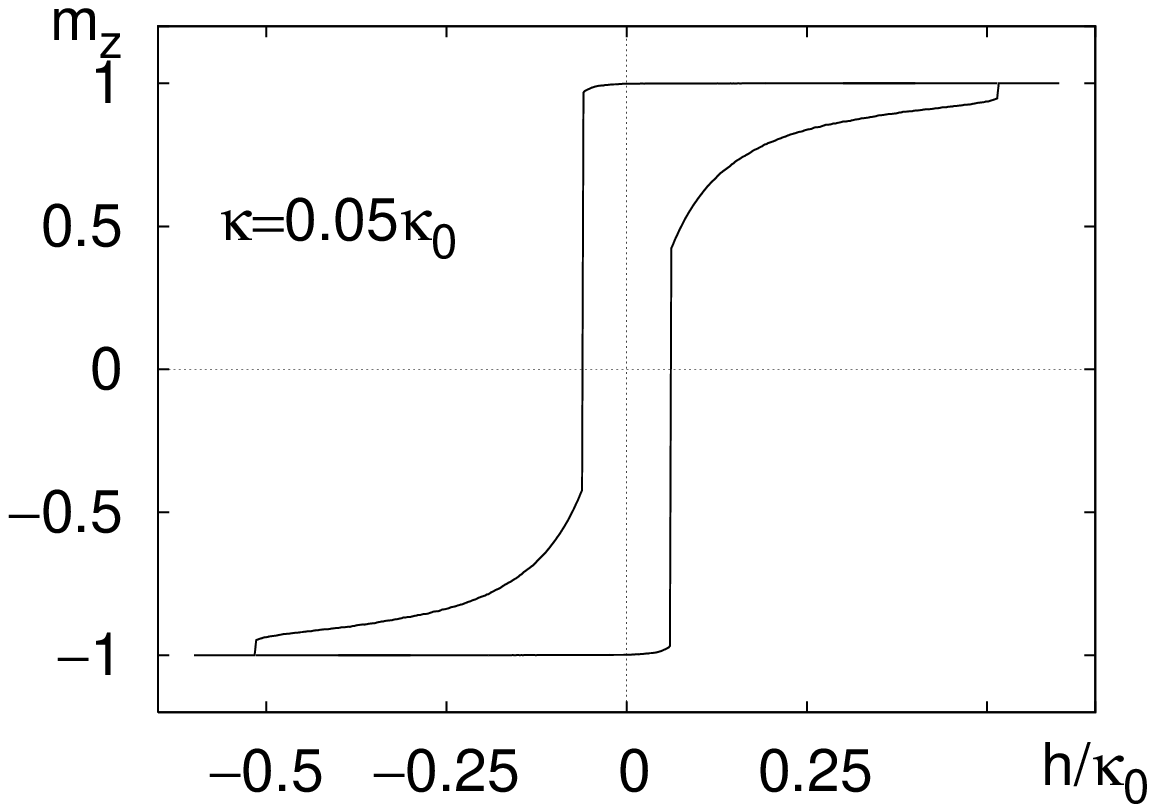}
                \includegraphics[width=0.95\columnwidth]{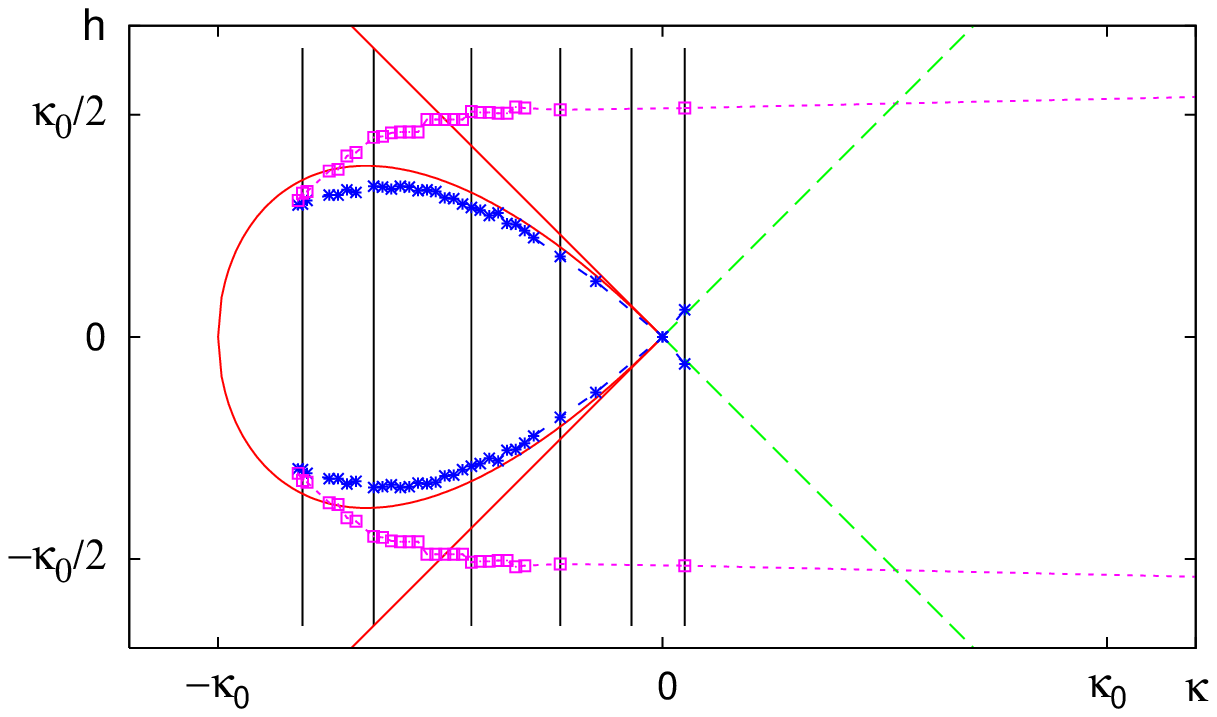}
            \end{center}
            \caption{(Color online)
            Numerical simulation of the out-of-plane hysteresis curves
            for several values of $\kappa/\kappa_0$. The locations of these sweeps
            in the phase diagram are shown by vertical lines
            in the bottom panel.}
            \label{all-hyst}
        \end{figure}

        In order to find the boundaries of stability of the striped
        phase, we performed a numerical simulation of the system
        described by Eqs.~(\ref{localE}) and (\ref{strayE}).  The
        simulation was conducted on a chain of 2048 magnetization
        vectors of unit length described by the angle $\theta(x)$.
        Periodic boundary conditions were applied.
        We used specific values for the exchange length and
        thickness of $5$ and $3$ lattice spacings, respectively.
        However, based on the scaling arguments described in Section
        \ref{scaling}, we expect the results to be universal for
        thin films. Eq.~(\ref{localE}) was minimized using a
        relaxation method,\cite{numrecipes} where $h$
        was replaced by $h_{\mathrm{eff}}(x)=h-\int\mathrm{d}x'\,
        V_1(x-x') \, \cos{\theta(x')}$ to account for the dipolar stray field.
        This $h_{\mathrm{eff}}$ was recalculated after a number of iterations of the relaxation
        method. The process continued until the Lagrange equation,
        \begin{equation}
            \lambda^2(\mathrm{d}^2\theta/\mathrm{d}x^2)=\kappa\cos\theta\sin\theta+h_{\mathrm{eff}}\sin\theta,
            \label{lagrange}
        \end{equation}
        was satisfied at each point. Random
        noise was then added to the system and $h$ was incremented.
        In this way, the algorithm moved the state of the system
        along the local minimum of energy. Sweeps of the
        magnetization were conducted from positive to negative
        values of $h$ and the average magnetization was recorded.
        The hysteresis loops shown in
        Fig.~\ref{all-hyst} were produced from
        these single sweeps by rotating the data points to produce
        the upward sweep and superimposing it on the downward one.
        Discontinuous jumps in the magnetization mark the system's
        entry into and exit from the striped phase. In what follows,
        we interpret the resulting phase diagram (Fig.~\ref{phase diagram}) for
        various values of the effective anisotropy.

        An important feature of phase diagram revealed by the
        numerical simulation is the existence of liquid-gas like
        critical points at $\kappa_c\approx -0.82\kappa_0$, $h_c\approx\pm0.3\kappa_0$
        (Fig.~\ref{phase diagram}). To
        the right of this point, a distinction can be made between a
        striped phase, with large modulation around a small average
        magnetization, and a spin-density wave phase, which
        has a small modulation around a larger average
        magnetization. Boundaries of metastability extend from
        the critical points, surrounding regions of coexistence
        between the striped and SDW phases. To the left of $\kappa_c$
         the two phases merge into one and
        magnetization curves proceed in a reversible manner. It is
        important to note that for $\kappa_c<\kappa<0$, the
        second-order phase transition line out of the canted phase
        is very close to the first-order phase transition line into the
        striped phase. As such, this second-order transition may be
        difficult to detect as it is hidden by the nearby first-order jump.

        In the remainder of this section, we describe the results of
        a sweep of the applied magnetic field beginning in high
        positive field normal to the sample and moving to a high
        negative field. The results of such a sweep depend on the
        anisotropy constant $\kappa/\kappa_0$ describing the system.

        For strong enough in-plane anisotropy, $\kappa<-\kappa_0$, the striped phase is
        never seen. A sweep of magnetic field from high positive to high
        negative fields would find a completely reversible magnetization
        curve, with the uniform upward phase beginning to cant at
        $h=-\kappa$ and following the field smoothly and linearly. The
        magnetization will be $-h/\kappa$ until the field is decreased to the
        $h=\kappa$ line, from which point onward the sample is downwardly
        polarized.

        For anisotropy values only slightly greater than the RPT value
        $\kappa=-\kappa_0$, a modulation of the magnetization will appear
        in low fields. The magnetization curve will still be reversible,
        but no longer exactly linear. Once the magnetic field crosses
        the curve $h=-\kappa\sqrt{1+\kappa/\kappa_0}$ from
        above, a small modulation out of the canted phase develops continuously.
        That modulation fades again at the lower curve
        $h=\kappa\sqrt{1+\kappa/\kappa_0}$, and the system
        proceeds as before from the canted to the uniform down phase.

        For anisotropy values greater than $\kappa_c$, the
        system shows history dependence in the magnetization
        curve.  At a slightly smaller value of $h$ than
        $h=-\kappa\sqrt{1+\kappa/\kappa_0}$, while the
        amplitude of the spin density wave is still on the order
        of $10^{-3}$, the system undergoes a
        first-order transition from a state with small modulation
        and large average magnetization to one with small average
        magnetization and a large amplitude modulation. This stripe
        phase remains as the field decreases until finally the
        system undergoes a first-order transition in which the
        periodic modulation disappears. The boundary of stability of the
        striped phase crosses $h=\kappa$, the line beyond which
        the downwardly magnetized state is stable, so that for
        more negative values of $\kappa$, the striped phase decays to a
        canted phase, while for more positive $\kappa$ values,
        the system enters the downwardly polarized state. For example, in
        Fig.~\ref{all-hyst}, the $\kappa=-0.81\kappa_0$ and
        $\kappa=-0.65\kappa_0$ magnetization curves show the striped phase
        decaying into the canted phase, while the other loops show a
        decay directly into the fully polarized phase.

        For a small $\kappa>0$, as we conduct a downward sweep of the field
        beginning at large positive values, the system undergoes a first-order
        phase transition out of the upwardly polarized state at $h=-\kappa$.
        There is no canted or spin-wave phase after the transition, but the
        simulations indicate that there is an intervening stable state
        with non-constant magnetization before the system decays into
        the downwardly polarized state. The limitations of the
        simulation become evident here. The periodic boundary conditions
        imposed in the simulation do not allow us to distinguish between
        a periodic state with a long period and a state with an isolated
        region of unfavored magnetization (soliton). Only a small number
        of solitons appeared within the period forced by the simulation.
        The considerations of Sections \ref{zero-trans} and \ref{collapse}
        indicate that for large enough $\kappa$ the isolated soliton state
        is stable while the striped state is not. Whether this
        holds at small $\kappa>0$ has not been determined.

        The non-uniform state resulting from the first-order
        phase transition out of the upwardly polarized state
        persists until it vanishes in a first-order transition
        to the downward state (at $h\approx-0.5\kappa_0$ in the simulation). If one were to reverse
        course and increase the field again before reaching this
        second transition, the upward regions that had been unfavored would grow slowly
        rather than the system returning immediately to the totally
        upwardly polarized phase. This behavior accounts for the ``fading
        contrast" seen in experiments with Co/Pt
        multilayers.\cite{Cheng06}

        If $\kappa>0$ is large enough, we expect that the transition will
        proceed directly from an upwardly polarized state to a
        downwardly polarized state at $h=-\kappa$. For very large
        $\kappa$, neither a striped nor an isolated-soliton phase can form
        from the decay of the upwardly polarized state, as neither will be stable
        in the high fields required for the uniform state to become unstable.
        (See Sections \ref{zero-trans} and \ref{collapse}).

        There is a caveat on this expectation, however. If
        topologically non-trivial domains of reversed
        magnetization form during the transition out of the
        upwardly polarized phase, it may be that they
        are stable even in these large fields. A topologically
        non-trivial domain in our case is one for which the
        magnetization angle $\theta$ at the left end of the domain
        differs by a non-zero multiple of $2\pi$ from the
        magnetization angle at the right end. It is not unreasonable
        to expect the formation of such domains: if two downward
        regions form during the transition
        by rotating in opposite directions away from $\theta=0$ then
        the upwardly polarized region between them
        will be topologically non-trivial. Since these downward
        regions may be initially well separated, their rotation directions are
        essentially independent, and so a non-trivial upward domain
        will form between them roughly half the time. These non-trivial
        domains are the one-dimensional analog of skyrmion-type domains
        in two dimensions.\cite{Polyakov75} Skyrmions may be the
        cause of the asymmetric domain nucleation observed in Co/Pt
        multilayers.\cite{Iunin06}

        The magnetization curve would look nearly the same in the
        case that topologically non-trivial solitons form as it would if
        they do not. The topologically non-trivial
        solitons would have small width when the field is large and
        so change the magnetization by a negligible amount. However,
        when the field is reversed after the transition, these
        soliton domains would grow. This would cause the system to proceed
        through the striped phase rather than staying downwardly
        polarized until $h=\kappa$. Hysteresis, then, would not be
        observed unless the applied field is strong enough to force
        the width of topological solitons down to the lattice scale,
        allowing them to decay.

    \subsection{$\kappa \gg \kappa_0$, the wide-stripe approximation}

        When the out-of-plane anisotropy is strong, the
        magnetization pattern is expected to consist of long up- or
        downwardly pointing regions, separated by short regions in
        which the magnetization changes rapidly from one domain type
        to the other. We refer to such a configuration as a ``wide-stripe"
         phase.  (Note, however, that this ``phase" is
        continuously connected to the modulated phase at negative
        $\kappa$, and so not a truly distinct phase.)

        It is well known that an energy advantage can be gained
        over a purely uniform phase through such modulation. In
        fact, when the out-of-plane anisotropy is very strong, the
        zero field model differs little from the dipolar Ising model
        discussed by Garel and Doniach.\cite{Garel82} The two
        significant differences lie in the energy costs of the kinks
        and in topological considerations. The topological consequences
        of the model will be discussed in Section \ref{collapse}.

        The wide-stripe configuration can be thought of as a
        periodic array of Bloch domain walls. These
        walls can be treated as elementary objects with some
        internal energy cost and long range interactions with other
        walls through the stray dipolar field. The interaction
        energy of two walls is logarithmic in their separation. If
        we define the orientation of a wall as a vector pointing
        from the upwardly magnetized side of the wall to the
        downwardly magnetized side, we find that the magnetic
        interaction is attractive for walls with the same
        orientation and repulsive for walls of opposite orientation.
        For a periodic structure, walls always have orientation
        opposite that of their nearest neighbors. It
        is this nearest neighbor repulsion caused by the stray magnetic
        field that allows for the stability of a striped phase despite
        the cost of the walls in exchange energy and the exchange force
        acting between nearby walls.

        These considerations can be used to derive an analytic
        expression for the magnetization curve in the wide-stripe phase:
        \begin{equation}
            \langle m_z\rangle=
        \frac{2}{\pi}\arcsin\left(\frac{\tilde{h}\pi^2}{4\sqrt{\tilde{\kappa}}}
        \exp\left(\frac{\pi}{2}\sqrt{\tilde{\kappa}}\right)\right),
        \end{equation}
        where we have used the rescaled variables
        \begin{equation}
        \tilde{\kappa} = \frac{\kappa}{\kappa_0}, \quad
        \tilde{h} = \frac{h}{\kappa_0}. \quad
        \end{equation}
        Note that this magnetization
        curve is consistent with the scaling property described in Section
        \ref{scaling}. Details of the derivation, including an
        analytic expression for the stripe period, can be found in
        Appendix \ref{ap-wide stripes}.

        \subsubsection{Zero-soliton-density transition}\label{zero-trans}
            As the external magnetic field is varied, the
            wide-stripe phase becomes dominated by the regions of
            magnetization favored by the magnetic field. For a
            strong enough magnetic field the striped phase
            has only narrow, widely separated regions of the
            unfavored magnetization (solitons). As
            Eq.~(\ref{stable period}) shows, the period of
            the wide-stripe phase tends to infinity along the curve
            $\tilde{h}=-(4/\pi^2)\sqrt{\tilde{\kappa}}\exp(-\pi\sqrt{\tilde{\kappa}}/2)$.            That is, the density of the unfavored solitons reaches zero.
            For fields above this curve, there is no stable
            structure with evenly spaced domains of unfavored
            magnetization. Note that the field at which the
            solitons are expelled from the system decreases exponentially with
            $\kappa$. For large $\kappa$ the region in which stripes
            exist is extremely narrow. As the field varies from down
            to up across this region, a system that is nearly all
            down with a few isolated upward solitons will move quickly
            through the striped phase to a state that is nearly all
            up with a few isolated downward solitons.

        \subsubsection{The collapse of isolated solitons}\label{collapse}

            In a perfectly isotropic sample, the field drives all the
            regions of unfavored magnetization to the edge of the
            system during the zero-soliton-density transition
            described above. A real sample, however, may have pinning
            centers where regions of opposing magnetization would be
            localized. These isolated solitons will persist into a
            much higher field, and their collapse is dependent on
            their topological character.

            The edges of such regions interact with each other through
            two distinct forces. There is a long-range repulsion
            between them due to the dipolar stray field and a
            short-range exchange interaction that may be attractive or
            repulsive depending on the topology of the region. In
            addition, there is a force from the applied magnetic field
            that acts to increase or decrease the width of the soliton
            based on its polarization. If the exchange force is
            attractive, the boundary of stability of the isolated
            soliton phase is found at the field that will squeeze the
            soliton boundaries enough that the exchange force takes
            over and the solitons collapse. If the exchange force is
            repulsive, the soliton must be forced to a width smaller
            than the lattice spacing in order to collapse, so that a
            phase difference of $2\pi$ between neighboring sites may
            be ignored as physically meaningless.

            The exchange force between two domain walls a distance
            $w$ apart
            is
            \begin{equation}\label{ex-force}
                \tilde{F}_\mathrm{ex}(\tilde{w}) \sim \pm
                8\tilde{\kappa}\exp{\left(-\sqrt{\tilde{\kappa}}
                \tilde{w}\right)}
                \mbox{ as } \sqrt{\tilde{\kappa}}\tilde{w} \to \infty,
            \end{equation}
            where $\tilde{w} = \sqrt{\kappa_0} w/\lambda$ (see
            Appendix \ref{exchange}). The force is attractive if the
            kinks form a non-topological soliton and repulsive if the
            soliton is topological.

            To determine the field necessary to collapse a
            non-topological soliton, we take the Zeeman energy
            of the soliton and the energy of the stray
            field to be
            \begin{equation}
                \tilde{E}(\tilde{w})= \mathrm{const} + 2\tilde{h}\tilde{w}
                -\frac{8}{\pi}\ln\tilde{w}
            \end{equation}
            for a single downward soliton of width $w$ [see
            Eq.~(\ref{singsolE})]. We use here the
            $\sqrt{\tilde{\kappa}}\tilde{w}\gg 1$ approximation, as
            we expect the collapse width of a non-topological
            soliton to be large compared to the
            size of the walls that border the soliton.

            As the field $\tilde{h}$ increases,
            the soliton is squeezed:
            \begin{equation}\label{squeeze}
            2\tilde{h}=\frac{8}{\pi\tilde{w}}
            - 8 \tilde{\kappa}
                \exp{\left(-\sqrt{\tilde{\kappa}} \tilde{w}\right)}.
            \end{equation}
            Note that as $\tilde{w}$ shrinks, the restoring force on the RHS of
            Eq.~(\ref{squeeze}) increases at first and then reaches a
            maximum at
            \begin{equation}
                \tilde{w} \sim \frac{\ln{\tilde{\kappa}}}{2\sqrt{\tilde{\kappa}}}
                 \quad \mbox{ as } \quad  \tilde{\kappa} \to \infty.
            \end{equation}
            A further increase in the field leads to a collapse of the soliton,
            as the restoring force can no longer balance the force due to the field.
            This justifies the approximation of large width, since
            for the collapse width
            $\sqrt{\tilde{\kappa}}\tilde{w}\sim (1/2)\ln{\tilde{\kappa}}\gg
            1$. Hence, in a field
            \begin{equation}
                \tilde{h} \sim \frac{8\sqrt{\tilde{\kappa}}}{\pi\ln{\tilde{\kappa}}}
                 \quad \mbox{ as } \quad  \tilde{\kappa} \to \infty,
            \end{equation}
            the (non-topological) soliton will collapse.

            If, however, the soliton consists of a full $2\pi$ rotation
            of the magnetization, the restoring force never reaches a
            maximum. While the dipolar contribution to the force effectively
            disappears as the soliton width decreases to the order of the domain
            wall width, the repulsive exchange force increases without bound.
            The soliton can then only collapse
            when its width
            \begin{equation}
                w=\frac{\tilde{w}\lambda}{\sqrt{\kappa_0}}=\frac{2\lambda}{\sqrt{\kappa+\abs{h}}}~{\rm
                arcsinh}\left(\sqrt{1+\frac{\kappa}{\abs{h}}}\right),
            \end{equation}
            derived in Appendix \ref{exchange},
            reaches the lattice scale.

\section{Stripes near an edge}\label{edge}

    In this section, we consider the orientation of stripes near a
    system edge. Although we begin with a variational solution with
    a sinusoidal modulation, the final result is applicable to a
    general profile for the magnetization. We find that stripes
    oriented with their domain walls perpendicular to the edge are
    energetically favored over all other orientations except
    possibly that with stripes {\em exactly} parallel to the edge.
    For simplicity, we consider only the case of no applied field.
    The results are entirely similar when a field is applied. In
    particular, there is no change to Eq. (\ref{orientation E}),
    below. We compare these results with recent experiments on thermally
    evaporated Ni films.

    We use the trial solution
    \begin{equation}
        m_z=a\sin{(\mathbf{q \cdot x} -\beta)}.
    \end{equation}
    Here $a$ is the amplitude of oscillations; the stripe wavevector
    $\mathbf q = (q\cos{\alpha}, \, q\sin{\alpha})$ has a fixed length
    $q$; the angle in the plane $\alpha$ ranges from $0$ for
    stripes normal to the edge to $\pi/2$ for
    stripes parallel to the edge; the phase $\beta$ will be important only
    for stripes parallel to the edge. In order to represent an edge at
    $y=0$, we include a a step function $\Theta(y)$ in all $y$
    integrals.

    The energy associated with the presence of an edge comes from two
    sources.  First, stripes at the edge generate a stray magnetic field.
    Second, depending on the sign of the effective anisotropy $\kappa$,
    the system can lower the energy by having an extra node or
    antinode near the edge.  (The latter works only for stripes parallel
    to the edge.)

    The long-range nature of the dipolar forces requires a certain amount of
    care in evaluating the edge energy.  The value of the dipolar
    energy is sensitive to three length scales: the film thickness $t$, the
    width $L_y$, and the stripe period $2\pi/q$.  On a computational level,
    keeping $t$ and $L_y$ finite is required to avoid ultraviolet and
    infrared divergences.  Fortunately, the {\em difference} in the energies
    between stripes with different orientations (parameterized by the angle
    $\alpha$) is insensitive to these length scales, as long as
    $\alpha \neq \pi/2$.  This simplifies the computation greatly.
    Details of the calculation can be found in Appendix
    \ref{ap-edgecalc}.

\begin{figure}[htbp]
        \begin{center}
            \includegraphics[width=0.9\columnwidth]{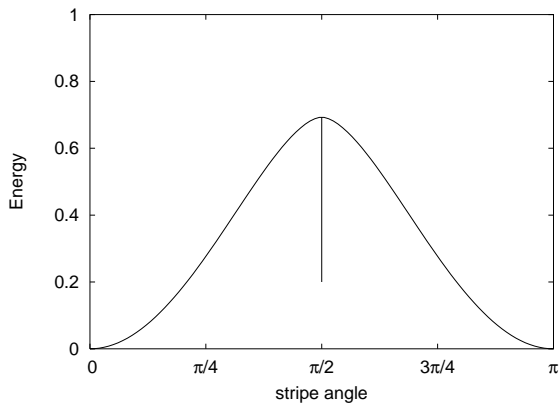}
        \end{center}
        \caption{
            Energy as a function of the angle the stripe wavevector
            makes with the system edge.  Energy here is measured in units of
            $(a^2/8\pi)\mu_0 M^2 t^2 L_x$.}
            \label{fig-edge-energy}
    \end{figure}

    For $\alpha\neq\pi/2$, the energy difference
    \begin{equation}
        \Delta E=E(\alpha,q)-E(0,q)
    \end{equation}
    has no ultraviolet or infrared divergences, and is
    in fact independent of $q$ (Fig.~\ref{fig-edge-energy}):
    \begin{equation}\label{orientation E}
        \frac{\Delta E(\alpha)}{\mu_0 M^2 t^2 L_x}=
        \frac{a^2}{8\pi} [\sin{\alpha}\ln{(1+\sin{\alpha})}
                    +(1-\sin{\alpha})\ln{\cos{\alpha}}].
    \end{equation}

    If $\alpha=\pi/2$, the stripes are oriented exactly
    parallel to the edge. In that case there is a cutoff dependent
    term in the stray field energy difference proportional to
    \begin{equation} \label{crest term}
        -\frac{a^2}{8\pi} \ln{(qt)}\cos{2\beta}.
    \end{equation}
    As long as the wavelength of the mode under consideration is
    larger than the thickness, the logarithm is negative. This term
    is therefore minimized when $\beta=\pi/2$ and so
    acts to attract the crests of the spin wave to the system edge.

    There is also an additional term associated with the local part of the free
    energy that appears only when $\alpha=\pi/2$. It is proportional to
    \begin{equation}\label{node term}
        -\frac{a^2}{8}\left(\lambda^2q^2+\kappa\right)\frac{\sin{(2\beta)}}{qt}.
    \end{equation}
    If the coefficient in brackets is positive (negative), this
    term acts to minimize the number of nodes (crests) of the spin wave
    that are present in the system, since each node (crest) has a
    cost in the exchange and anisotropy energy. The competition
    between the terms (\ref{crest term}) and (\ref{node term}) will
    determine the phase of a spin-density wave parallel to the edge.

    The addition of the phase-dependent terms above may mean that
    stripes parallel to the edge end up being the global minimum of
    energy. However, this minimum takes
    the form of a downward spike in the energy that occurs at an angle that would
    otherwise be a maximum of the energy (see
    Fig.~\ref{fig-edge-energy}). Even if the state with stripes
    parallel to the edge is the global minimum of energy, a low
    temperature system that begins with stripes at a random
    orientation is more likely to evolve toward the (metastable) minimum at
    $\alpha=0$.

    Since the angular dependence of the energy for $\alpha\neq \pi/2$ is
    independent of the wavenumber $q$, this result holds for any
    stripe profile that may be made up of higher harmonics.  In that
    case, $a^2/2$ should be interpreted as the average value
    $\langle \cos^2{\theta} \rangle$ of the striped state.
    The result is also independent of any of
    the material parameters because it is the energy of the stray
    field that causes the effect.

\begin{figure}
        \begin{center}
            \includegraphics[width=0.85\columnwidth]{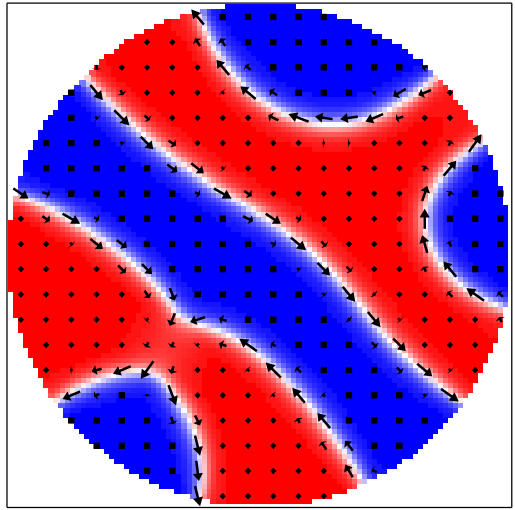}
        \end{center}
        \caption{(Color online)   A stationary
            configuration obtained from a random initial state in a
            disk of thickness 14 nm and diameter 400 nm.
            Magnetization length $M = 1.4 \times 10^6$ A/m, exchange
            constant $A = 3.3 \times 10^{-11}$ J/m, exchange length
            $\lambda = 5.2$ nm, easy-axis anisotropy $K = 1.5\times
            10^6$ J/m$^3$ yield $\kappa = 0.22$ and $\kappa_0 =
            0.45$.  Magnetization points up in the red (light gray)
            regions and down in the blue (dark gray) regions.
            Numerical simulation using OOMMF.\cite{oommf}}
        \label{fig-disk}
    \end{figure}

    This result is consistent with observations made in thermally
    evaporated thin nickel films\cite{Lee06} and numerical simulations
    (Fig. \ref{fig-disk}) that
    indicate that walls meet the edges of the film at a
    $90^\circ$ angle.

\section{Discussion}

    We have examined the properties of thin ferromagnetic films with a
    strong easy-axis anisotropy $K \approx K_0 = \mu_0 M^2/2$ favoring
    the out-of-plane component of the magnetization, with Co/Pt films as
    a prototype.  We have found a scaling property that applies in the
    limit where the film thickness is the shortest length scale in the
    problem.  In such a case the phase diagram in the field-anisotropy
    plane is universal if the applied field $\mathbf H$ and effective
    anisotropy $K - K_0$ are properly rescaled.  The proper variables
    are $\tilde{\mathbf h} = \mathbf H/(\kappa_0 M)$ and $\tilde{\kappa}
    = (K-K_0)/(\kappa_0 K_0)$. The parameter $\kappa_0 = (t/4\lambda)^2$
    is determined by the film thickness $t$ and the exchange length
    $\lambda = \sqrt{2A/\mu_0 M^2}$.

    The universal phase diagram in the case of an applied field normal
    to the plane was determined through a combination of analytical and
    numerical methods focussed on uniform states and on states with a
    one-dimensional modulation of the magnetization, such as magnetic
    stripes observed near the reorientation phase transition ($K \approx
    K_0$).  In addition to fully magnetized and canted uniform states,
    at least two non-uniform magnetic states were found: a spin-density
    wave and a striped phase. These two states are found to coexist in
    parts of the phase diagram (like a gas and a liquid).  The
    coexistence of various phases (e.g., stripes and SDW or stripes and
    canted) explains the rich hysteretic behavior of magnetization
    observed in these films.  We have determined the boundaries of
    stability and metastability of these phases and obtained
    out-of-plane magnetization curves $M_\perp(H_\perp)$, which have
    universal shapes determined by the rescaled anisotropy
    $\tilde{\kappa}$.

    In addition to developing the zero-temperature phase diagram
    near the reorientation phase transition, we have expanded on
    previous work in the case of large out-of-plane anisotropy by
    finding analytic expressions for the stripe period and
    magnetization as functions of the applied field. We have shown
    that the range of field values for which a structure of evenly
    spaced stripes is stable is exponentially small for large
    anisotropy. The period of such a structure tends to infinity
    at a small value of the applied field. Stripes are unlikely
    to be found, then, at large anisotropies.

    We have investigated the behavior of stripes near the film edge and
    found that the energy of the dipolar stray field is minimized when
    the stripes are perpendicular to the edge, as recently observed in
    Ni films. At the same time we find that a state with stripes
    parallel to the edge provides an opportunity to lower other terms in
    the free energy, e.g. the magnetic anisotropy, by registering the
    nodes or antinodes of magnetization at the system edge.

    The behavior of the striped phase in an out-of-plane magnetic field
    depends on the topology of the stripes, i.e. on the direction of
    rotation of the magnetization.  If in the absence of an applied
    field the colatitude angle $\theta$ oscillates between (nearly) 0
    and $\pi$, the topology is trivial.  As the strength of the applied
    field increases, the stripes of the ``wrong'' orientation first
    shrink and then disappear leaving behind a uniform state.  However,
    if the domain walls delineating a ``wrong'' domain wind in the same
    direction (say from 0 to $\pi$ and from $\pi$ to $2\pi$), this
    domain has a nontrivial topology.  As its wall are pushed together,
    they feel strong repulsion mediated by exchange.  As a result, such
    a domain will not decay until its size reaches a microscopic scale
    (presumably on the order of a lattice constant), at which point the
    phase increase from 0 to $2\pi$ can be repaired through a phase
    slip.  These nontrivial domains are the one-dimensional analog of
    the skyrmion,\cite{Polyakov75} a texture with a nonzero O(3) winding
    number. Skyrmions may play a role in the magnetization reversal of
    thin films with high anisotropy.\cite{Iunin06}

\section{Acknowledgements}

    We thank C.-L. Chien, S. H. Lee, N. Markovic, V. I. Nikitenko, 
    and F. Q. Zhu for stimulating this work and for sharing unpublished 
    data. This work was supported in part by NSF Grant DMR-0520491 and 
    by the JHU Theoretical Interdisciplinary Physics and Astronomy Center.

\appendix

\section{Periodic structure of wide stripes}\label{ap-wide stripes}
    When walls between uniform regions are well
    separated we can treat them as nearly free defects
    interacting with one another mostly via their stray magnetic
    fields. The stray fields of other walls are thus
    substantially weak and can be neglected when determining the
    structure of a well-isolated wall. Furthermore, its own
    stray field can be neglected as well: the energy associated
    with the stray field is $\mathcal{O}(t^2)$, whereas
    all the other energies are $\mathcal{O}(t)$. Finally, when
    the anisotropy highly favors out-of-plane magnetization, the
    characteristic width of the domain walls is small compared to
    that of the stripes, so that the
    energy to be gained by deforming the domain walls in the
    presence of an external magnetic field is far outstripped by
    the energy to be gained by moving them. Hence the external
    field may be neglected as well when determining the domain
    wall structure.

    It is convenient
    to use dimensionless variables for the effective anisotropy,
    field, coordinate, and wavenumber:
    \begin{equation}
        \tilde{\kappa} = \frac{\kappa}{\kappa_0}, \quad
        \tilde{h} = \frac{h}{\kappa_0}, \quad
        \tilde{x} = \frac{x\sqrt{\kappa_0}}{\lambda}, \quad
        \tilde{k} = \frac{k\lambda}{\sqrt{\kappa_0}}.
    \end{equation}
    In terms of these variables, the energy (\ref{localE})-(\ref{strayE})
    can be expressed as
    \begin{equation}
        \frac{E_\mathrm{local}}{\mu_0 M^2 L_y t^2}
        =\frac{1}{4}
        \int\mathrm d \tilde{x}\left[
        \frac{1}{2} \left(\frac{\mathrm d \theta}{\mathrm d \tilde{x}}\right)^2
        -\frac{\tilde{\kappa}}{2}\cos^2\theta - \tilde{h} \cos{\theta}\right]
        \label{rlocalE}
    \end{equation}
    and
    \begin{equation}
        \frac{E_\mathrm{stray}}{\mu_0 M^2 L_y t^2}=
        -\frac{1}{4}\int\frac{\mathrm{d}\tilde{k}}{2\pi}\abs{\tilde{k}}
        \abs{{\widetilde{m}_z}({\tilde{k}})}^2,
        \label{rstrayE}
    \end{equation}
    where we use the Fourier transform of the out-of-plane magnetization
    \begin{equation}
        \widetilde{m}_z(\tilde{k})=\int
        \mathrm{d}\tilde{x} \, e^{i\tilde{k}\tilde{x}}\cos\theta(\tilde{x})
    \end{equation}
    We neglect for the moment the effects of the external and stray
    magnetic field in order to find the internal structure of a domain
    wall.  A domain wall that interpolates between
    \mbox{$\cos\theta=-1$} and $\cos\theta=1$ and minimizes the sum of
    exchange and effective anisotropy terms in the energy above obeys:
    \begin{equation}
        \theta'^2=\tilde{\kappa}\sin^2\theta.
    \end{equation}
    The derivative here is with respect to $\tilde{x}$.
    This is solved by $m_z=\cos\theta=\pm \tanh(\sqrt{\tilde{\kappa}}\tilde{x})$.
    The internal energy of each such wall is $\mu_0
    M^2t^2L_y\sqrt{\tilde{\kappa}}/2$, not
    counting the interaction energy due to the stray field.

    When calculating the stray field energy,
    it is easier to work with the $\tilde{x}$-derivative
     of the magnetization, and
    use the form:
    \begin{equation}\label{stray int}
        \frac{E_\mathrm{stray}}{\mu_0 M^2 L_y t^2}=
        -\frac{1}{4}\int\frac{\mathrm{d}\tilde{k}}{2\pi}
        \frac{\abs{\widetilde{m'}_z({\tilde{k}})}^2}{\abs{\tilde{k}}}.
    \end{equation}

    If the walls have small spatial extent relative to the
    distance between them, a periodic
    structure can be approximated as a sum of alternating upward
    and downward kinks. As such, we describe a periodic structure with
    period $l=\tilde{l}\lambda/\sqrt{\kappa_0}$ and upward length
    $w=\tilde{w}\lambda/\sqrt{\kappa_0}$ in each period by
    the variational solution
    \begin{equation}\label{periodic structure}
        m_z'(\tilde{x})=\int \mathrm{d} u\,
        \Delta(\tilde{w},\tilde{l},\tilde{x}-u)
        \frac{\mathrm{d}}{\mathrm{d}u}(\tanh(\sqrt{\tilde{\kappa}}u)),
    \end{equation}
    where
    \begin{equation}
        \Delta(\tilde{w},\tilde{l},u)=
        \sum_{n=-\infty}^\infty
        \delta(u-n\tilde{l})-\delta(u-n\tilde{l}-\tilde{w}).
    \end{equation}
    By substituting (\ref{periodic structure}) into
    (\ref{stray int}), we arrive at
    \begin{eqnarray}
        \frac{E_\mathrm{stray}}{\mu_0 M^2 V\kappa_0}&=&
        -\frac{16\pi^3}{\tilde{\kappa}\tilde{l}^3}
        \sum_{j=0}^\infty j\frac{\sin^2(\pi j
        \tilde{w}/\tilde{l})}{\sinh^2(\pi^2
        j/\sqrt{\tilde{\kappa}}\tilde{l})}\nonumber\\
        &=&\frac{4(2\pi/\tilde{l})^3}{\tilde{\kappa}}
        \frac{\mathrm{d}}{\mathrm{d}\rho}\sum_{j=0}^\infty\sum_{n=1}^\infty
        \sin^2(\pi j\tilde{w}/\tilde{l}) e^{-2\rho j n},\nonumber\\
    \end{eqnarray}
    where $\rho=\pi^2/\sqrt{\tilde{\kappa}}\tilde{l}$ and
    $V=tL_xL_y$ is the volume of the film.

    Summing over $j$ leads
    to an expression
    that can be approximated by an integral over $n$ if
    $\rho \ll 1$.  We thus obtain
    \begin{equation}\label{period stray}
        \frac{E_\mathrm{stray}}{\mu_0 M^2 V\kappa_0}=
        -\frac{4}{\pi\tilde{l}}\left[\ln{(1+f)} 
         + \frac{2f}{1+f}\right],
    \end{equation}
    where
    \begin{equation}
        f(\tilde{w},\tilde{l})=\frac{\tilde{l}^2\tilde{\kappa}}{\pi^4}\sin^2\left(\frac{\pi
        \tilde{w}}{\tilde{l}}\right).
    \end{equation}
    The total internal energy of the kinks in this structure 
    (including exchange and anisotropy) is 
    \begin{equation}\label{kinksE}
        \frac{E_{\mathrm{kinks}}}{\mu_0 M^2V\kappa_0}=
        \frac{4\sqrt{\tilde{\kappa}}}{\tilde{l}},
    \end{equation}
    since there are two kinks in each period. The final contribution
    to the energy is the interaction
    with the applied magnetic field, described by the energy
    density
    \begin{equation}\label{period h}
        \frac{E_{\mathrm{ext.\, field}}}{\mu_0 M^2V\kappa_0}=
        -\frac{2\tilde{h}\tilde{w}}{\tilde{l}}.
    \end{equation}

    Minimizing the sum of these three terms with respect to $\tilde{l}$ and $\tilde{w}$
    leads to the following expressions for the equilibrium period and upward width of
    the stripes [for $\tilde{l}\sqrt{\tilde{\kappa}}\sin(\pi \tilde{w}/\tilde{l})\gg
    1$]:
    \begin{equation}\label{width in l}
        \cos\left(\frac{\pi
        \tilde{w}}{\tilde{l}}\right)=-\frac{\tilde{h}\pi^2}{4\sqrt{\tilde{\kappa}}}
        \exp\left(\frac{\pi}{2}\sqrt{\tilde{\kappa}}\right)
    \end{equation}
    and
    \begin{equation}\label{stable period}
        \tilde{l}\sqrt{\tilde{\kappa}}\sin(\pi
        \tilde{w}/\tilde{l})=\pi^2
        \exp\left(\frac{\pi}{2}\sqrt{\tilde{\kappa}}\right).
    \end{equation}
    Note that Eq.~(\ref{stable period}) justifies the approximation
    that $\tilde{l}\sqrt{\tilde{\kappa}}\sin(\pi \tilde{w}/\tilde{l})\gg
    1$, since $\sqrt{\tilde{\kappa}}$ is large in this region of the
    phase diagram.

    These equations can be solved to give:
    \begin{equation}
        \tilde{w}=\frac{2\arccos(-\tilde{h}/\tilde{h}_0)}
                       {\pi\sqrt{\tilde{h}_0^2-\tilde{h}^2}}
    \end{equation}
    and
    \begin{equation}
        \tilde{l}=\frac{2}{\sqrt{\tilde{h}_0^2-\tilde{h}^2}}
    \end{equation}
    with
    $\tilde{h}_0=-(4/\pi^2)\sqrt{\tilde{\kappa}}\exp(-\pi\sqrt{\tilde{\kappa}}/2)$.

    For $\tilde{h}=0$ the above equations reduce to $\tilde{w}=\tilde{l}/2$
    (i.e., no net
    magnetization), and a stable period of
    \begin{equation}
        l=\frac{\lambda \tilde{l}}{\sqrt{\kappa_0}}=
        \frac{\pi^2\lambda}{\sqrt{\kappa}}
        \exp\left(\frac{\pi}{2}\sqrt{\frac{\kappa}{\kappa_0}}\right).
    \end{equation}

    Since the average magnetization \mbox{$\langle m_z\rangle= 2\tilde{w}/\tilde{l}-1$},
    Eq.~(\ref{width in l}) can be rewritten to give the magnetization curve for
    large $\kappa$:
    \begin{equation}
        \langle m_z\rangle=
        \frac{2}{\pi}\arcsin\left(\frac{\tilde{h}\pi^2}{4\sqrt{\tilde{\kappa}}}
        \exp\left(\frac{\pi}{2}\sqrt{\tilde{\kappa}}\right)\right).
    \end{equation}

    Further, note that as
    \mbox{$\tilde{h}\to -\tilde{h}_0$}, the period
    $\tilde{l}$ tends to infinity while the upward width stays
    finite \mbox{[$\tilde{w}\to
    (\pi/4\sqrt{\tilde{\kappa}})\exp(\pi\sqrt{\tilde{\kappa}}/2)$]}. For
    larger fields there is no stable periodic structure of this
    type.

    If, instead of a periodic structure, we are interested in the
    energy of a single soliton, we take $l \to L_x \gg w$ in
    Eqs.~(\ref{period stray})-(\ref{period h}), so that
    $f(\tilde{w},\tilde{l})\approx \tilde{\kappa}\tilde{w}^2/\pi^2$ leading to
    \begin{eqnarray}\label{singsolE}
        \frac{4E}{t^2L_y\mu_0 M^2}&\approx&
        4\sqrt{\tilde{\kappa}} - 2\tilde{h}\tilde{w}
        -\frac{4}{\pi}\left[\ln\left(1+\frac{\tilde{\kappa}\tilde{w}^2}{\pi^2}
        \right)+\frac{2
        \tilde{\kappa}\tilde{w}^2}{\pi^2+\tilde{\kappa}\tilde{w}^2}\right]
            \nonumber\\&\approx&
            \mathrm{const} - 2\tilde{h}\tilde{w}-\frac{8}{\pi}\ln{\tilde{w}}
    \end{eqnarray}
    for an upward soliton of width $w$ when $\tilde{\kappa}\tilde{w}^2 \gg 1$.

\section{Solitons and the exchange force between
domain walls}\label{exchange}

    The energy of a single soliton given above by Eq.~(\ref{singsolE}) does not
    take into account the exchange interaction between the domain walls
    bounding the soliton. This interaction will become important as
    the walls move closer together.

    If we wish to find the effective force due to the exchange
    interaction, we must first find stable soliton solutions to the
    Lagrange equation associated with the energy (\ref{rlocalE}). We
    will ignore the effects of the stray magnetic field (in the limit of
    large $\kappa$ with $t\ll\lambda/\sqrt{\kappa-\abs{h}}$). Since the
    force on the walls due to the magnetic field is known, the stable width of a
    soliton can be used to find the effective exchange force between the walls that must
    be acting to oppose the field.

    We will apply the boundary conditions $\theta(\pm \infty)=\pi$
    and $\theta'(\pm \infty)=0$, thus describing an upward soliton
    in a downwardly polarized background. We obtain from (\ref{rlocalE}) that
    \begin{equation}\label{diffeq-sol}
        \frac{\theta'^2}{2}=\frac{1}{2}\tilde{\kappa}\sin^2\theta-\tilde{h}(1+\cos\theta),
    \end{equation}
    where $\tilde{\kappa}=\kappa/\kappa_0$, $\tilde{h}=h/\kappa_0$, and
    $\tilde{x}=x\sqrt{\kappa_0}/\lambda$, as in Eqs.~(\ref{rlocalE})-(\ref{rstrayE}).

    We expect different solutions to this equation for $\tilde{h}>0$ and
    $\tilde{h}<0$. If $\tilde{h}>0$, then the soliton is favored by
    the field, and the background is unfavored, held in place only by the
    out-of-plane anisotropy. In order to balance the force of the
    field, the exchange interaction in this case will attract the
    walls of the soliton to one another.

    The corresponding solution to Eq.~(\ref{diffeq-sol}) at $\tilde{h}>0$ is
    \begin{equation}\label{sol-non-topo}
        \cos\theta=-1+\frac{2(1-\tilde{h}/\tilde{\kappa})}{1
    +(\tilde{h}/\tilde{\kappa})\sinh^2(k\tilde{x})},
    \end{equation}
    where $k=\sqrt{\tilde{\kappa}-\tilde{h}}$.

    Note that
    $\cos\theta(0)=1-2\tilde{h}/\tilde{\kappa}$ is the maximum value
    of $\cos\theta$ in this solution. Nowhere is there full upward
    polarization, so this solution describes a non-topological
    soliton, in which there is no net rotation of the magnetization
    between the ends of the system.

    If, on the other hand, $\tilde{h}<0$, then the background is
    favored by the field and the soliton is not. The exchange force
    between the walls of the soliton must be repulsive in order to
    balance the force of the field squeezing the walls together.
    This is accomplished by a topological soliton solution, in which
    the magnetization rotates by a full $2\pi$ between the ends of
    the system. The solution to Eq.~(\ref{diffeq-sol}) in this case ($\tilde{h}<0$)
    is
    \begin{equation}\label{sol-topo}
        \cos\theta=-1+\frac{2(1-\tilde{h}/\tilde{\kappa})}{1-(\tilde{h}/\tilde{\kappa})\cosh^2(k\tilde{x})}.
    \end{equation}
    Again, $k=\sqrt{\tilde{\kappa}-\tilde{h}}$.

    Note that in this case,
    $\cos\theta(0)=1$ for any value of $\tilde{h}$. The $2\pi$ rotation
    of the magnetization forces the magnetization to point upward in
    the center of the soliton in any field. This is the reason for
    the repulsive exchange force. If the soliton is squeezed, the
    magnetization is forced to move from down to up and back in a
    shorter distance.

    If we set $\cos\theta=0$ in Eq.~(\ref{sol-non-topo}) to find the
    locations $\pm \tilde{x}_0$ of the kinks bounding the non-topological
    soliton, we obtain
    \begin{equation}
        \tilde{h}=\frac{\tilde{\kappa}}{1+\cosh^2(\tilde{x}_0\sqrt{\tilde{\kappa}-\tilde{h}})}\approx
        4\tilde{\kappa}\exp\left(-2\sqrt{\tilde{\kappa}}\tilde{x}_0\right)
    \end{equation}
    for large $\tilde{\kappa}$. Since the force of the field is
    $2\tilde{h}$ acting to separate the kinks, the (rescaled) exchange force in
    the non-topological case is
    \begin{equation}
        \tilde{F}_\mathrm{ex} = -8\tilde{\kappa}
        \exp\left(-\sqrt{\tilde{\kappa}}\tilde{w}\right).
    \end{equation}
    Note that $\tilde{w}=2\tilde{x}_0$, since the kinks are at
    $\tilde{x}=\pm \tilde{x}_0$.

    Similarly, if we set $\cos\theta=0$ in Eq.~(\ref{sol-topo}), we
    obtain
    \begin{equation}\label{pretopowidth}
        -\tilde{h}=\frac{\tilde{\kappa}}{-1+\sinh^2(\tilde{x}_0\sqrt{\tilde{\kappa}-\tilde{h}})}\approx
        4\tilde{\kappa}\exp\left(-2\sqrt{\tilde{\kappa}}\tilde{x}_0\right)
    \end{equation}
    for large $\tilde{\kappa}$ and moderate width. Since the force of
    the field is now acting to compress the kinks, the (rescaled) exchange force in
    the case in which the soliton is topological is $\tilde{F}_\mathrm{ex}=+8\tilde{\kappa}\exp\left(-\sqrt{\tilde{\kappa}}\tilde{w}\right)$, a
    repulsive force.

    If the field is very high, the kinks in the topological case will be
    forced close together, so that the large $\sqrt{\tilde{\kappa}}\tilde{w}$
    approximation is no longer valid. The soliton will not collapse,
    however, until the width [obtained by solving
    Eq.~(\ref{pretopowidth}) for $\tilde{w}=2\tilde{x}_0$]
    \begin{equation}\label{topowidth}
        w=\frac{\tilde{w}\lambda}{\sqrt{\kappa_0}}=\frac{2\lambda}{\sqrt{\kappa+\abs{h}}}~{\rm
        arcsinh}\left(\sqrt{1+\frac{\kappa}{\abs{h}}}\right)
    \end{equation}
    is on the order of the lattice spacing.

\section{Energy of stripes near an edge}
\label{ap-edgecalc}

    The energy of a two-dimensional thin-film system with magnetic field
    oriented normal to the plane of the sample and no bulk charge is
    \begin{eqnarray}\label{2DE}
        \frac{E}{\mu_0 M^2 t}&=&\int\mathrm{d}^2x\left[
        \frac{\lambda^2|\nabla
        m_z|^2}{2(1-m_z^2)}-\frac{\kappa}{2}m_z^2-hm_z\right]\nonumber\\
        &&-\frac{t}{4}\int
        \frac{\mathrm{d}^2k}{(2\pi)^2}\abs{k}\abs{{m_z}({\mathbf{k}})}^2 .
    \end{eqnarray}
    Here $m_z({\mathbf{k}})$ represents the
    two-dimensional Fourier transform of $m_z$.
    In order to include the effects of a system edge, we will include a step
    function $\Theta(y)$ as a factor in all the $y$-integrals that appear.  For
    simplicity we discuss the case of zero applied field, $h=0$, only.

    We use a trial solution with a single wave number and
    propagation direction. We will be comparing the energy of solutions with
    different values of the angle between the propagation direction
    and the system edge. We label this angle $\alpha$.
    Our trial solution is
    \begin{equation}\label{trial}
        m_z = a\sin{\left(\mathbf{q \cdot x}-\beta\right)},
    \end{equation}
    where $a$ is the modulation amplitude, \mbox{$\mathbf q =
    (q\cos{\alpha}, \, q\sin{\alpha})$}, and $\alpha$ ranges from $0$
    for stripes running perpendicular to the edge to $\pi/2$
    for stripes running parallel to the edge. In order to evaluate
    the energy, we use the form
    \begin{equation}\label{step function}
        \Theta(y)=\int\frac{\mathrm{d}p}{2\pi}\frac{i e^{ipy}}{p+i\epsilon}
    \end{equation}
    for the step function in order to regularize the infrared
    divergence. Here $L_y=1/\epsilon$ is the extent of the system
    in the $y$-direction. Where necessary, we also use the inverse of
    the sample thickness as an ultraviolet cutoff
    in momentum space integrals.

   \subsection{$\alpha \neq \pi/2$}

    As long as stripes are not parallel to the edge, trial solutions
    with different values of $\beta$ are related to one another by a
    translation in the $x$ direction. Since the integration in
    Eq.~(\ref{2DE}) is carried out over all $x$, the energy is
    independent of the phase $\beta$.  Further, the total
    contribution of the local terms in the energy is the same for
    any orientation other than $\alpha= \pi/2$. The dependence of
    energy on the orientation then reflects the effect of the
    dipolar stray field alone.

    By inserting the trial solution (\ref{trial}) into the stray
    field term in Eq.~(\ref{2DE}) using the form (\ref{step function})
    for the step function, we obtain directly that the contribution
    of the stray field to the energy is:
    \begin{eqnarray}\label{divergent stray E}
        \frac{E_{\mathrm{stray}}(\alpha,q)}{E_0}=
        -\frac{a^2}{16}\int\frac{\mathrm{d}p}{2\pi} \Big(
        \frac{\sqrt{q^2 + 2qp \sin{\alpha} + p^2}}{p^2+\epsilon^2}
        \nonumber\\
        +\frac{\sqrt{q^2 - 2qp \sin{\alpha} + p^2}}{p^2+\epsilon^2}\Big),
    \end{eqnarray}
    where we introduce a characteristic energy scale
    \begin{equation}
        E_0=\mu_0 M^2 t^2 L_x.
    \end{equation}
    This integral requires both infrared and ultraviolet cutoffs.
    The inverse thickness $1/t$ serves to cut off the ultraviolet divergence
    in the integral and $\epsilon=1/L_y$ the infrared. However,
    the difference $E_{\mathrm{stray}}(\alpha,q) - E_{\mathrm{stray}}(0,q)$ is not
    sensitive to these cutoffs and can be evaluated in the limits $t \to 0$
    and $1/\epsilon = L_y \to \infty$.  Remarkably, the difference is also
    independent of the wavenumber $q$. By subtracting $E_{\mathrm{stray}}(0,q)/E_0$
    from Eq.~(\ref{divergent stray E}) and performing the $p$ integration,
    we obtain
    \begin{equation}
        \frac{E_{\mathrm{stray}}(\alpha,q) - E_{\mathrm{stray}}(0,q)}{E_0}
        = \frac{a^2}{8\pi} f(\alpha),
    \end{equation}
    where the function
    \begin{equation}
    f(\alpha) = \sin{\alpha}\ln{(1+\sin{\alpha})}
    +(1-\sin{\alpha})\ln{(\cos{\alpha})}
    \end{equation}
    has a minimum at $\alpha=0$, when the stripes are normal to the edge.

    \subsection{$\alpha = \pi/2$}

    When stripes are parallel to the edge, the energy also depends on the
    phase $\beta$.  To leading order in $t$ and $\epsilon$,
    \begin{eqnarray}
        \frac{E(\pi/2,q)-E(0,q)}{E_0} &=&\frac{a^2}{8\pi}\Big[
         -\frac{\pi(\kappa + \lambda^2 q^2)\sin(2\beta)}{qt}
        \nonumber\\
        &&+\ln(2)-\ln(qt)\cos(2\beta)\Big].
    \end{eqnarray}
    In addition to the energy of the stray field, this expression reflects
    the energy of anisotropy and exchange, both of which are sensitive to
    the positions of nodes relative to the edge.

    The overall energy dependence, including a dip at $\alpha = \pi/2$,
    is shown in Fig.~\ref{fig-edge-energy}. Depending on the parameters of
    the film, the global minimum may be either at $\alpha = 0$ or at
    $\alpha = \pi/2$.  However, even if the global minimum is at $\pi/2$,
    the system may not easily find that configuration and remain in the
    metastable state with the stripes normal to the edge ($\alpha=0$).

\end{document}